\documentclass{ws-p9-75x6-50}
\begin{document}

\title{Striped quantum Hall phases}

\author{Felix von Oppen,$^{(1)}$ Bertrand I.\ Halperin,$^{(2)}$ and Ady Stern
$^{(3)}$}

\address{$^{(1)}$ Institut f\"ur Theoretische Physik, Universit\"at 
zu K\"oln, Z\"ulpicher Str.\ 77, 50937 K\"oln, Germany\\
$^{(2)}$ Physics Department, Harvard University, Cambridge, Massachusetts 
02139,USA\\
$^{(3)}$ Department of Condensed Matter Physics, The Weizmann Institute of 
Science, 76100 Rehovot, Israel}

\maketitle

\abstracts{Recent experiments seem to confirm predictions that
  interactions lead to charge density wave ground states in higher
  Landau levels.  These new ``correlated'' ground states of the
  quantum Hall system manifest themselves for example in a strongly
  anisotropic resistivity tensor. We give a brief introduction and
  overview of this new and emerging field.}

\section{Introduction}

While a large number of fractional quantized Hall states have been
discovered in the lowest Landau level, such states become increasingly
rare in higher Landau levels. Few quantized Hall states have been
observed in the first Landau level (Landau level filling factor
$2\le\nu<4$) and no such states have so far been found for filling
factors $\nu\ge4$. Correspondingly, the role of electron-electron
interactions in this filling factor range has long remained almost
uncharted territory.

This is now rapidly changing, following recent experiments by Lilly
{\it et al.}\cite{Lilly} and by Du {\it et al.}\cite{Du} on extremely
clean two-dimensional electron systems (2DES) for Landau level (LL)
filling factor $\nu>4$. The central observation is that the
resistivity becomes strongly anisotropic close to half filling of the
topmost Landau level in this filling factor range.  This is believed
to be a signature of a novel Coulomb-induced charge density wave
ground state whose existence had been predicted by Fogler {\it et
  al.}\cite{Fogler} and by Moessner and Chalker.\cite{Moessner} Unlike
the Laughlin states,\cite{Laughlin} these states can be partially
understood within the Hartree-Fock (HF) approximation. In fact, it was
already suggested at the end of the '70s by Fukuyama {\it et
  al.}\cite{Fukuyama} that the HF ground state of the lowest Landau
level is a charge density wave (CDW).  While in the lowest Landau
level, the CDW ultimately turned out to be preempted by the Laughlin
states, it seems to be more favorable in higher Landau levels.

Different ground states have been predicted depending on filling
factor.\cite{Fogler,Moessner} Near half filling of higher Landau
levels ($\nu\simeq N+1/2$), the HF ground state should be a
unidirectional charge density wave (UCDW). In this state, the filling
factor alternates between stripes of filling factor $N$ and $N+1$.
The period of the density modulation is of the order of the cyclotron
radius $R_c$.  Further away from half filling one expects a so-called
bubble phase, in which clusters of the minority filling factor ($N$ or
$N+1$) order on a triangular lattice with characteristic length $R_c$.
Very close to integer filling factors, a Wigner crystal phase should
form.

The strongly anisotropic transport properties\cite{Lilly,Du,Shayegan}
near half filling of higher Landau levels are believed to be associated
with the unidirectional charge density wave phase.  In this paper we
briefly review some of the recent work on these states.  The
experimental results are reviewed in Sec.\ \ref{sec-experiment}.
Sec.\ \ref{theory} focuses on theoretical developments.  In Sec.\ 
\ref{sec-theory} we sketch early theoretical work which suggested
that there exists an instability towards CDW formation. It
is well known that Hartree-Fock calculations tend to overestimate the
degree of ordering.  Some insights into this can be gained from an
analogy with two-dimensional liquid crystal phases which we review in
Sec.\ \ref{sec-liquid}. A focus of recent theoretical work are the
transport properties of the UCDW states, as reviewed in Sec.\ 
\ref{sec-transport}. It has been shown under rather general
assumptions that the conductivity tensor should satisfy non-trivial
relations, which are independent of microscopic
parameters.\cite{MacDonald,Oppen-cdw} Finally we summarize and
conclude in Sec.\ \ref{sec-conclusions} by mentioning some open
problems.

\section{Experiment}
\label{sec-experiment}

The most prominent observation
\cite{Lilly,Du,Shayegan,Pan,Lilly2,Eisenstein} is the development of
large anisotropies in the resistivity close to half filling of the
topmost Landau level. Observation of this effect requires extremely
clean samples and very low temperatures ($T<150$mK). The anisotropy
develops even though the sample behaves essentially isotropically at
very large and very low magnetic fields.  The case for a new
``correlated'' ground state in this filling factor range is further
strengthened by the observation that the width of the peak in the
resistivity around half filling does not decrease with decreasing
temperature.  This is very different from what one would expect for
the integer quantum Hall effect (QHE) plateau transition.

To date, the principal experimental features of these novel states
are:

(a) The anisotropic resistivity has so far been observed near half
filling of the topmost LL in the filling factor range $4\le\nu\le12$.
The anisotropy is largest for $\nu=9/2$ and decreases monotonically
with increasing LL index $N$.\cite{Lilly,Du}

(b) The principal axes of the anisotropic resistivity tensor seem to
be consistently oriented along certain crystallographic axes of the
GaAs crystal.  The high-resistance direction is always along the
$1{\bar1}0$ direction while the low-resistance direction is aligned
with the $110$ direction.\cite{Lilly,Du}

(c) The resistances in the two principal directions differ by up to
several orders of magnitude in van-der-Pauw
measurements.\cite{Lilly,Du} However, van-der-Pauw measurements
significantly overestimate the anisotropy of the resistivity tensor
(assuming that a local resistivity tensor is an appropriate
description).\cite{Simon,Lilly-comment} This is associated with the
detailed current distribution in the device.  Accordingly, Hall-bar
measurements which should be free of this problem\footnote{However,
  they currently measure the two diagonal resistivities on different
  samples.} show a much smaller but still very significant anisotropy
of the order of five.\cite{Lilly}

(d) The anisotropy usually appears more stable against temperature for
filling factors corresponding to the lower spin component of each
LL.\cite{Lilly,Du}

(e) A striking set of experiments\cite{Pan,Lilly2} revealed that even
weak in-plane magnetic fields $B_\parallel$ can have a strong effect
on the anisotropic states (and on the even denominator fractional QHE
state at $\nu=5/2$.)  Applying the in-plane field in the
low-resistance direction leads to an interchange of easy and hard
directions for $B_\parallel$ of the order of $0.5$T. In-plane fields
in the high-resistance direction affect the measured resistivities in
the upper spin component of each LL only very weakly, but suppress the
anisotropy in the lower spin component.

(f) Transport in the high-resistance direction is strongly non-linear
with the differential resistance rising with increasing bias current.
The change in resistivity with applied current is smooth, which would
seem inconsistent with a depinning transition of the UCDW.  The
non-linearity is more pronounced in the lower spin component of each
LL.\cite{Lilly}

(g) Intriguing reentrant {\it integer} QHE states have been found near
quarter and three quarter filling of the uppermost LL.\cite{Cooper}
The Hall resistance in these states is quantized at the value of the
closest integer plateau. Current-voltage characteristics in this
filling factor regime exhibit discontinuous and hysteretic behavior.
It has been suggested that this may be related to depinning of a CDW
state.\cite{Cooper}

\section{Theory}
\label{theory}

\subsection{Hartree-Fock calculations and numerical exact-diagonalization 
studies}
\label{sec-theory}

It turns out that essential features of the CDW ground states can be
understood within the Hartree-Fock approximation (HFA). Most
calculations assume that one can project to a single (partially
filled) Landau level. This is natural for the lowest LL. For higher
LLs, this relies on an effective interaction derived by Aleiner and
Glazman\cite{Aleiner} which includes the effects of screening by the
lower (filled) LLs. This effective interaction turns out to be
sufficiently weak compared to the Landau level spacing so that
projection to a single LL should be a reasonable starting point.

The zero-temperature HF equations within the single-LL approximation
have been studied in detail by Fogler {\it et al.}\cite{Fogler} In
particular, these authors compared various possible CDW ground states.
In the vicinity of half filling of the topmost LL ($\nu\simeq N+1/2$)
they find that a unidirectional CDW state whose period is of the order
of the cyclotron radius is most favorable. In this state,
one-dimensional stripes of filling factor $N$ alternate with stripes
of filling factor $N+1$. The analysis of this state is simplified by
the fact that the exact HF eigenfunctions are still the usual Landau
gauge wave functions for electrons in a magnetic field.  All Landau
gauge states with centers in the filling factor $N+1$ range are
occupied, while those wavefunctions with centers in the filling factor
$N$ region remain empty. Clearly, the relative modulation in the
filling factor is larger than that of the electron density due to the
finite width of the wavefunctions.

While CDW formation leads to a cost in Hartree energy due to the
associated charge density modulation, this energy cost is more than
offset by the gain in exchange energy due to the closer packing of the
electrons. It turns out that this mechanism is particularly effective
in higher LLs where the (Hermite-polynomial) wavefunctions have
zeroes. In this case, there are wavevectors of the charge density wave
for which the associated {\it charge} density modulation is
particularly small. Essentially, this happen when a state centered in
the middle of a filling factor $N+1$ region has its first side maximum
in the filling factor $N$ region. It is for this reason that the CDW
states are more favorable in higher LLs than in the lowest LL. 

The bulk of this paper will be concerned with the unidirectional CDW.
Further away from half filling, Fogler {\it et al.}\cite{Fogler}
predict a triangular CDW, termed bubble phase, consisting of clusters
of minority filling factor ordering in a background of majority
filling factor on a triangular lattice. Again, cluster size and period
of the lattice are given by the cyclotron radius $R_c$.  Very close to
integer filling factors, the electrons (or holes) in the topmost
Landau level are predicted to form a triangular Wigner crystal.

Moessner and Chalker\cite{Moessner} arrived at similar conclusions by
deriving Landau theories for various CDW states and comparing their
free energy. Following Fukuyama {et al.},\cite{Fukuyama} these authors
expand the free energy in powers of the appropriate order parameter.
This procedure is justified in the vicinity of the Hartree-Fock
transition temperature $T_c^{hf}$.  In addition, they show by
diagrammatic arguments that the Hartree-Fock approximation becomes
exact for the uniform phase in the limit of high Landau levels.

These conclusions have been partly supported by numerical exact
diagonalization studies.\cite{Rezayi} In these calculations, systems
of up to 12 electrons on a torus are diagonalized exactly and strong
peaks indicating CDW ordering have been found in the wave vector 
dependence of the static density susceptibility and the equal-time
density-density correlation function.

Motivated by the dramatic effect of in-plane magnetic fields
$B_\parallel$ in experiment, two groups\cite{Jungwirth,Stanescu}
extended the Hartree-Fock calculations to include the finite thickness
of the 2DES and the resulting orbital effects of $B_\parallel$.  While
these calculations suggest that the influence of $B_\parallel$ is
sensitive to sample details, the perhaps more realistic of the two
calculations\cite{Jungwirth} shows that in-plane fields can rotate the
stripe pattern in the appropriate manner. These calculations do not
explain the experimentally observed dependences on the spin of the LL.

\subsection{Analogy with liquid crystal systems}
\label{sec-liquid}

Hartree-Fock calculations tend to overestimate the degree of ordering
of a system. The influence of quantum and thermal fluctuations in the
present system can be assessed to some degree by an analogy with
two-dimensional (2d) liquid crystals.\cite{Fradkin} The analogy is
based on the fact that the UCDW shares its symmetry with 2d smectic
liquid crystals. The appropriate elastic variable of both systems is
the phase $u$ of the density oscillations, $\delta\rho({\bf
  r})=\rho_0\cos(q_0 x-u({\bf r}))$.  In terms of $u$, the
long-wavelength elastic free energy in the absence of forces tending
to align the stripes reads\cite{Nelson}
\begin{equation}
\label{free-energy}
  F={A\over 2}\int d{\bf r}\left\{\left(\partial u\over\partial
    x\right)^2 +\lambda^2\left(\partial^2u\over\partial
    y^2\right)^2\right\},
\end{equation}
where $A$ is an elastic constant and $\lambda$ a length which is
presumably of the order of the UCDW period.  The absence of a term
involving $(\partial u/\partial y)^2$ is a consequence of the global
rotation symmetry of the system. (Note that $u=ay$ only leads to a
global rotation of the CDW to linear order in $a$ and thus leaves the
free energy unchanged.)

It is an immediate consequence of this elastic free energy that
dislocations will cost only a finite energy.\cite{Nelson} Thus there
will be a finite density of dislocations at any non-zero temperature
and the stripes should be broken down to stripe segments.  This is
expected to destroy translational long-range order but preserve
quasi-long-range orientational order of the remaining stripe segments.
Thus, the smectic order predicted by the HFA is preserved only at zero
temperature. At non-zero temperatures, the system is analogous in terms
of symmetries to a 2d nematic liquid crystal.

As the temperature increases the system should undergo a
Kosterlitz-Thouless transition from the nematic phase to an isotropic
phase in which the orientational order of the stripe segments
disappears. Nevertheless, short-range stripe ordering should still
persist up to the presumably much higher Hartree-Fock transition
temperature $T_c^{hf}$.

MacDonald and Fisher\cite{MacDonald} have recently studied the
stability of the UCDW against quantum fluctuations in the framework of
the elastic theory and found that quantum fluctuations are not strong
enough to destroy the smectic order predicted by the HFA.

While much can be learned about the influence of quantum and thermal
fluctuations from this analogy, much less is known about the influence
of disorder on the UCDW.\cite{Radzihovsky} One may expect that
transport properties of the striped phases should be affected by even
small amounts of disorder on the substrate, which will pin the stripe
positions at low temperatures. Disorder should also lead to a finite
density of dislocations, even at zero temperature. Moreover, since the
forces aligning the stripes are believed to be very weak, steps or
other large-scale features of the GaAs-AlGaAs interface may lead to
large regions where the stripes are oriented differently from the
average preferred direction.

\subsection{Transport properties}
\label{sec-transport}

An important problem is a quantitative understanding of transport in
the striped phases. It had been previously noted in various
contexts\cite{modulation} that an {\it externally imposed} density
modulation can lead to strongly anisotropic transport in the presence
of a magnetic field.  While this effect is indeed related to the
development of anisotropic transport in higher Landau levels, none of
the developed formalisms apply directly to the present system.

\begin{figure}
\centerline{\psfig{figure=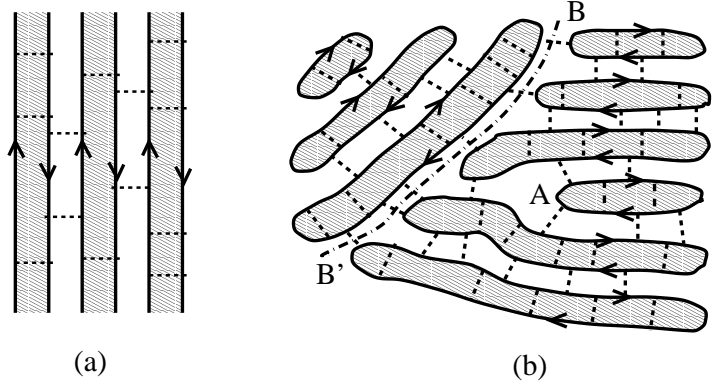,width=5.0cm,angle=270}}
\caption{(a) Ideal and (b) realistic stripe structures. 
  Shaded and unshaded regions represent stripes with filling factors
  $N + 1$ and $N$, respectively. Arrows indicate the direction of
  electron flow on edge states.  The dashed lines are scattering
  centers allowing electrons to tunnel between neighboring edge
  states.  The realistic stripe structure contains a dislocation (A)
  and a large angle grain boundary (B-B$^\prime$).}
\label{stripe-structure}
\end{figure}

We start with the idealized case of a {\it pinned} unidirectional CDW
near $\nu=N+1/2$, first studied by MacDonald and
Fisher\cite{MacDonald} (cf.\ Fig.\ \ref{stripe-structure}a). The
stripes are taken parallel to the $y$ axis and the CDW has period $a$.
Current can be carried by the {\it internal} edge channels belonging
to the $(N+1)$st LL which travel along each stripe edge, with velocity
$v_F$ and density of states $N(0)=1/hv_F$. Impurities lead to
scattering between these edge channels.  The scattering rates should
decrease rapidly with increasing distance between edges.  Thus, we
include only scattering between neighboring edges and denote the
scattering rate across electron stripes (stripes of filling factor
$N+1$) as $1/\tau_e$ and across hole stripes (stripes of filling
factor $N$) as $1/\tau_h$.  Each scattering event acts as
backscattering with regard to the $y$ direction and a single
random-walk step in the $x$ direction.  In addition, we assume that
quantum interference does not play an important role.  In this case,
we can obtain the conductivity tensor via the Einstein relation from
the classical diffusion constants.

The density of excitations on the up-moving and down-moving edge
channels of the $i$-th electron stripe, denoted by $P^+_t(y,i)$ and
$P^-_t(y,i)$ respectively, satisfy the rate equations
\begin{eqnarray}
   (\partial_t+v_F\partial_y)P^+_t(i)&=&-\left({1\over\tau_e}+{1\over\tau_h}
    \right)P^+_t(i)
    +{1\over\tau_h}P_t^-(i-1)+{1\over\tau_e}P_t^-(i) 
    \label{rate1}\\
   (\partial_t-v_F\partial_y)P^-_t(i)&=&-\left({1\over\tau_e}+{1\over\tau_h}
    \right)P^-_t(i)
    +{1\over\tau_e}P_t^+(i)+{1\over\tau_h}P_t^+(i+1) .
\label{rate2}
\end{eqnarray}
Acting on Eq.\ (\ref{rate1}) with $\partial_t-v_F\partial_y
+(1/\tau_h+1/\tau_e)$, inserting Eq.\ (\ref{rate2}), and dropping
second-order time derivatives, one obtains the diffusion equation
\begin{eqnarray}
   \partial_t P^+_t(i)={v_F^2\over2}{\tau_e\tau_h\over\tau_e+\tau_h}
   \partial_y^2 P^+_t(i)
    +{1\over 2(\tau_e+\tau_h)}[P_t^+(i+1)-2P_t^+(i)+P_t^+(i-1)].
\label{diffusion}
\end{eqnarray}
The same equation is satisfied by $P^-_t(i)$. One can now read off the 
diffusion constants 
\begin{eqnarray}
   D_{xx}={a^2\over2}{1\over\tau_e+\tau_h} 
   \qquad\qquad
   D_{yy}={v_F^2\over2}{\tau_e\tau_h\over\tau_e+\tau_h}.
\end{eqnarray}
According to the Einstein relation, the conductivity is related to the
diffusion constants by $\sigma_{\alpha\alpha}=e^2(2/a)(1/hv_F)
D_{\alpha\alpha}$ and one obtains\cite{MacDonald}
\begin{eqnarray}
\label{conductivity-ideal}
   \sigma_{xx}={e^2\over h}{a\over v_F(\tau_e+\tau_h)} \qquad\qquad
   \sigma_{yy}={e^2\over h}{v_F\over a}{\tau_e\tau_h\over\tau_e+\tau_h}.
\end{eqnarray}

To obtain the Hall conductivity, we note that the $N$ completely
filled LL's contribute $Ne^2/h$. To find the contribution of the
partially filled topmost LL, we apply a chemical potential gradient in
the $x$ direction.  Assuming a chemical potential drop of $ev$ between
the two edges of an electron stripe, we have a Hall current
$j_y=(e^2/h)(v/a)$ in the $y$ direction and a diffusion current
$j_x=(e/\tau_e)(1/hv_F)ev$ in the $x$ direction. Comparing with
$j_x=\sigma_{xx}E_x$, we find $v=[\tau_e/(\tau_e+\tau_h)] E_x a$ and
therefore a Hall conductivity of\cite{MacDonald}
\begin{equation}
\label{hall-ideal}
  \sigma_{xy}={e^2\over h}\left(N+{\tau_e\over\tau_e+\tau_h}\right).
\end{equation}

Remarkably, these results make a number of predictions which are
independent of the microscopic parameters.\cite{MacDonald,Oppen-cdw}
Following MacDonald and Fisher\cite{MacDonald} we first focus on the
symmetric point $\tau_e=\tau_h\equiv\tau$. If one assumes
particle-hole symmetry in the partially filled LL, this would
correspond to half filling of the topmost Landau level, $\nu=N+1/2$.
In this case, we deduce from Eq.\ (\ref{conductivity-ideal}) that the
product of the diagonal conductivities takes on a universal
value\cite{MacDonald}
\begin{equation}
    \sigma_{xx}\sigma_{yy}=(e^2/2h)^2,
\label{product-rule}
\end{equation}
independent of the period, Fermi velocity, or the scattering rate.
Likewise, one finds that the Hall conductivity becomes independent of
the microscopic parameters, $\sigma_{xy}={e^2\over h}(N+1/2)$.
Hence, we can also rewrite the product rule in terms of the diagonal 
resistivities as\cite{MacDonald}
\begin{equation}
    \rho_{xx}\rho_{yy}=(h/e^2)^2{1\over [N^2+(N+1)^2]^2}.
\label{product-rule-rho}
\end{equation}
For the Hall resistivity at the symmetric point one finds 
$\rho_{xy}=(h/e^2)(2N+1)/[N^2+(N+1)^2]$. 

While the product of the diagonal resistivities is universal, the
anisotropy $\rho_{xx}/\rho_{yy}$ depends on the microscopic parameters,
\begin{equation}
\label{anisotropy}
  {\rho_{xx}\over\rho_{yy}}=\left(v_F\tau\over a\right)^2,
\end{equation}
and is given by the square of the ratio of the basic diffusion steps
in the $y$ and $x$ directions.

It is also interesting to note that if we assume that the diagonal
conductivity (or, equivalently, the scattering rates $1/\tau_e$ and
$1/\tau_h$) do not depend significantly on the Landau level index
$N$, say at the symmetric point, then the resistivities decrease with
increasing Landau level index roughly as $1/N^2$. This seems to be in
rather good agreement with the experimental results for the peak
heights in the high-resistance direction.

Initial results indicate that the product rule Eq.\ 
(\ref{product-rule-rho}) is reasonably well satisfied in
experiment.\cite{Eisenstein-priv} This may be surprising in view of
the expectation that the experimental samples should be quite far from
the perfect stripe ordering assumed here.  Thus, experiment raises the
question whether the product rule is not in fact valid much more
generally than indicated by this derivation. An additional question is
related to the neglect of quantum interference.  Since the
experimental anisotropy in the resistivity is about
five,\cite{Lilly,Simon} these results [cf.\ Eq.\ (\ref{anisotropy})] would
imply that the electrons hop between edges after traveling only a
distance of a few cyclotron radii along the edge.  For such a
situation, quantum interference effects should be important,
particularly since the experiments are performed at extremely low
temperatures. While it is currently not known how quantum interference
affects the validity of the product rule, it is certainly possible
that it leads to significant deviations.

Transport in more realistic stripe structures has been studied by von
Oppen {\it et al.}\cite{Oppen-cdw} The starting point is that the
conductivity tensor of the perfect stripe structure actually satisfies
an even more general relation, namely {\it the semicircle
  law}\cite{Oppen-cdw}
\begin{equation}
  \sigma_{xx}\sigma_{yy}+(\sigma_{xy}-\sigma_h^0)^2=(e^2/2h)^2,
\label{semicircle}
\end{equation}
with $\sigma_h^0={e^2\over h}(N+1/2)$, which {\it holds also away from
  the symmetric point.} This relation can be generalized to a wide
class of stripe structures. In particular, a semicircle law holds even
in the presence of topological and orientational defects such as
dislocations and grain boundaries if one assumes that the defects are
pinned by disorder. An example of such a more general stripe structure
is shown in Fig.\ \ref{stripe-structure}(b).

Strictly speaking, one finds that the macroscopic conductivity tensor
$\hat\sigma^*$ (relating the spatially averaged currents and fields)
satisfies the semicircle law\cite{Oppen-cdw}
\begin{equation}
  \sigma^*_1\sigma^*_2+(\sigma^*_h-\sigma_h^0)^2 = (e^2/2h)^2.
\label{semicircle*}
\end{equation}
Here, the macroscopic conductivity tensor
$\hat\sigma^*=\hat\sigma^*_d+\sigma_h^*\hat\epsilon$ is written as the
sum of its dissipative part $\hat\sigma_d^*$ and Hall component
$\sigma_h^*\hat\epsilon$ with $\hat\epsilon$ the totally antisymmetric
tensor.  $\sigma_1^*$ and $\sigma_2^*$ are the eigenvalues of the
(real symmetric) dissipative part $\hat\sigma_d$.

The product rule Eq.\ (\ref{product-rule}) is a special case of the
semicircle law for the symmetric point $\sigma_{xy}=(e^2/h)(N+1/2)$.
Thus, the more general validity of the semicircle law also implies the
same for the product rule, thereby explaining its agreement with
experiment. Moreover, this makes the experimental results consistent
with a picture where electrons hop between edges much more rarely,
while the anisotropy is reduced by the presence of defects such as
dislocations and grain boundaries.  In such a picture, neglecting
quantum interference may indeed be justified.\cite{Oppen-cdw}

The generalized semicircle law Eq.\ (\ref{semicircle*}) can be
supported by two different arguments.\cite{Oppen-cdw} The most general
derivation uses results obtained by Shimshoni and
Auerbach\cite{Shimshoni} for a model of the ``quantized Hall
insulator.'' In this argument, one maps the problem to a network of
puddles of filling factor $\nu=1$ in vacuum. Neglecting quantum
interference, assuming that the network is planar, and that the stripe
structure is pinned, it was shown by Shimshoni and Auerbach that the
corresponding (properly defined) Hall resistance equals
$R_{xy}=h/e^2$. Following the mapping from the stripe structure to the
puddle network in reverse, one finds that this implies the semicircle
law (\ref{semicircle*}).\cite{Oppen-cdw} The assumption that the
network is planar would for example be violated if one included
next-to-nearest neighbor hopping between edges.

Alternatively, one can also give a continuum argument\cite{Oppen-cdw}
for the semicircle law Eq.\ (\ref{semicircle*}) based on a duality
transformation first exploited by Dykhne and Ruzin.\cite{Dykhne} In
this argument, one assumes that the defect density is sufficiently low
so that one can define a local conductivity tensor which everywhere
satisfies the semicircle relation (\ref{semicircle}). The defects lead
to spatial variations of the scattering rates and of the principal
axes of the dissipative part of the conductivity tensor. Using a
duality transformation, involving new currents and fields which are
linear combinations of the original currents and fields, and choosing
the dual system to be the time reverse of the original one, it can
then be shown\cite{Oppen-cdw} that the macroscopic conductivity tensor
satisfies the semicircle law Eq.\ (\ref{semicircle*}).

Several authors observed that the internal edge modes should be
Luttinger liquids.\cite{Fradkin,MacDonald} It has been
argued\cite{MacDonald} that the stripe phase is unstable (possibly
towards an anisotropic Wigner crystal) due to backscattering, albeit
only below experimentally accessible temperatures. MacDonald and
Fisher\cite{MacDonald} have also suggested that the Luttinger liquid
behavior may explain the experimentally observed nonlinearities in the
diagonal resistivity. Within the transport theory sketched above,
Luttinger liquid correlations imply that the scattering rates
$1/\tau_e$ and $1/\tau_h$ decrease with increasing voltage.  Thus, the
anisotropy increases with increasing voltage (or applied current), as
was observed in experiment. However, there are not yet detailed
predictions for the full temperature and current dependences which
might be compared with experiment.

Fradkin {\it et al.}\cite{Fradkin2} have used the liquid-crystal
analogy to study the temperature dependence of the anisotropy for the
transition from the nematic to the isotropic phase.  They argue on the
basis of symmetry that close to $T_c$ the anisotropy should be
proportional to the order parameter of an xy model with director order
parameter in the presence of a small background anisotropy.  Using
Monte-Carlo results for this model they attempt to fit the
experimental anisotropy, using two free parameters. The validity of
the analysis is hard to assess, however, because it is unknown for
which range around $T_c$ the assumed proportionality holds.

\section{Conclusions and open questions}
\label{sec-conclusions}

The discovery of the anisotropic phases in higher Landau
levels\cite{Lilly,Du}, combined with the earlier theoretical
predictions of CDW states\cite{Fogler,Moessner} in this filling factor
range, has opened a new chapter of quantum Hall physics.  Initially,
the experimental evidence for charge density wave ordering has been
purely qualitative.  The product rule\cite{MacDonald} and the
semicircle law\cite{Oppen-cdw} discussed in Sec.\ \ref{sec-transport}
are the first predictions for the striped CDW phases which can be
tested {\it quantitatively} in experiment. Initial results seem to
indicate reasonable agreement.\cite{Eisenstein-priv}

It seems clear that the investigation of the new charge density wave
states in higher Landau levels has only started and much interesting
physics remains to be studied. Obvious open questions include the
following: The mechanism which causes the stripes to line up
preferentially with a particular axis of the GaAs substrate is not
well understood. The influence of disorder on the striped phases has
not been studied.  There is currently no explanation for the
observation that various quantities have a prominent dependence on
whether the Fermi energy is in the lower or upper spin component of a
LL. The influence of quantum interference on transport has not been
studied.  Filling factors away from half filling of the topmost LL
have received relatively little attention.

\section*{Acknowledgments}

We benefitted from discussions with J. Eisenstein, B. Huckestein, M.
Shayegan, S.  Simon, D. Shahar and H. Stormer.  The work was supported
in part by SFB 341, NSF grant DMR-94-16910, US-Israel BSF grant
98-354, DIP-BMBF grant, a Minerva foundation grant, and the Israeli 
Academy of Science.
%Initial phases of the work were carried out at the ITP, Santa Barbara,
%in August 1998, with support from NSF grant PHY94-07194.

\end{document}